\documentclass[prl,showpacs,twocolumn,preprintnumbers,psfig,amsmath,amssymb]{revtex4}

\usepackage{epsfig}
\usepackage{graphicx}
\usepackage{dcolumn}
\usepackage{bm}

\newcommand{\vf}{v_{\rm F}}

\begin{document}

\title{Midgap states and  charge inhomogeneities
in corrugated graphene}
\author{F. Guinea$^1$}
\affiliation{$^1$Instituto de Ciencia de Materiales de Madrid (CSIC),
Cantoblanco, Madrid 28049, Spain}
\author{M. I. Katsnelson$^2$}
\affiliation{$^2$Condensed Matter Theory, Institute for Molecules and
  Materials, Radboud University
  Nijmegen, Toernoooiveld 1, NL-6525 ED Nijmegen, The Netherlands}
\author{M. A. H. Vozmediano$^3$}
\affiliation{$^3$Instituto de Ciencia de Materiales de Madrid (CSIC),
Cantoblanco, Madrid 28049, Spain}

\begin{abstract}
We study the changes induced by the effective gauge field due to
ripples on the low energy electronic structure of graphene. We
show that zero energy Landau levels will form, associated to the
smooth deformation of the graphene layer, when the height corrugation, $h$,
and the length of the ripple, $l$, are such that $h^2 / ( l a ) \gtrsim 1$,
where $a$ is the lattice constant. The existence of localized levels  gives
rise to a 
large compressibility at zero energy, and to the enhancement of
instabilities arising from electron-electron interactions
including electronic phase separation. The combined effect of the
ripples and an external magnetic field breaks the valley symmetry
of graphene leading to the possibility of valley selection.
\end{abstract}

\maketitle

\section{ Introduction.}
The discovery of atomic thick graphene
layers\cite{Netal05b}, and the measurement of their novel
electronic properties\cite{GN07} have led to a great deal of
interest. The details of the structure of these layers is not
completely elucidated, although there is clear evidence that free
standing sheets are not flat\cite{Metal07}, and layers placed on
an insulating substrate seem also not to be flat, following the
corrugation of the susbtrate\cite{Setal07,ICCFW07}. The height and
width of the observed ripples in free standing samples are
consistent with models which take into account the tendency of
carbon ions to form different types of bonds\cite{FLK07}.

The intrinsic curvature of the graphene
sheets\cite{GGV92,GGV93b,MG06,CV07,JCV07}, the modulation of
hoppings by elastic strains\cite{SA02b,Metal06,M07}, and
hybridization between the $\pi$ and $\sigma$ bands induced by
curvature\cite{NK07} enter into the effective Dirac equation which
describes the low energy electron states as a deformation induced
gauge field. In the absence of scattering between the two
inequivalent valleys in the Brillouin Zone, these gauge fields
change the phases of coherent electrons, and can play a role
similar to that of an effective magnetic
field\cite{Metal06,MG06,Mcetal06,GTMHS07,WLSBH07,THGS07}, changing the
transport properties of the system in the presence of  low (real)
magnetic fields.

Single layer graphene also shows inhomogeneities in the electronic
distribution near half filling\cite{GN07,Metal07b}. These
inhomogeneities may be related to the existence of
ripples\cite{NK07,JCV07}.

In the present work, we study the changes induced by the effective
gauge field due to ripples on the low energy electronic structure
of graphene. These changes become important when the effective
magnetic length due to the ripples is comparable to the ripple
size. Then, zero energy Landau levels\cite{PGN06} can exist in
regions where the effective field changes slowly. These midgap
states are not suppressed by the off diagonal disorder associated
to the ripples\cite{Netal05}, and lead to a peak in the density of
states at zero energy. While clean graphene is a semimetal, with a
vanishing electronic compressibility at half filling, the
existence of ripples lead to a large compressibility at zero
energy, and to the enhancement of instabilities arising from
electron-electron interactions. Note that zero gap states are expected when
the flux of the effective magnetic field is larger than one in a given
region\cite{J84}. It has also been shown that a sufficiently strong random
gauge field leads to a divergent density of states at zero energy\cite{HD02}.

We present  estimates of the strength of the effective field
induced by ripples of the sizes observed experimentally, and
discuss simple models which illustrates the formation of Landau
levels as function of the deformation of the graphene layer. 
We also analyze the combined effects of ripples and a real magnetic
field, showing that the equivalence of the two valleys in the
Brillouin Zone of graphene is broken. We finally discuss possible
instabilities which can arise due to the enhanced electronic
compressibility, including the possibility of electronic phase
separation. The Appendix describes simple models where the phenomena discussed
in the paper can be studied analytically.

\section{Effective gauge field in single layer graphene.}
\subsection{Qualitative estimates.}
Fluctuations
of order $\Delta t$ of the hopping parameter $t$ around an hexagon
of the honeycomb lattice give rise to an effective flux, in units
of $e h / c$, through the hexagon of order $\Delta \Phi \sim
\Delta t / t$. If the hopping varies smoothly over a distance $l$,
then $\Delta t \approx \delta t ( a / l )$, where $\delta t$ is
the overall modulation of $t$, and $a$ is a distance comparable to
the lattice spacing. Then, the total flux in an region of area
$l^2$ is $\Phi \sim ( \delta t / t ) ( l / a )$. This number is
also an estimate of the number of Landau levels with $n=0$ within
that area. Finally, the magnetic length associated with the
effective field is $l_B \sim \sqrt{(
  t / \delta t ) ( l a )}$, and the separation between the $n=0$ and $n =
\pm 1$ Landau levels is $\sqrt{ ( t \delta t ) ( a / l )}$. If the origin of
the modulation $\delta t$ is solely due to the strains induced by a ripple of
height $h$ and length $l$, then $\delta t \approx \partial \log ( t ) /
\partial \log ( d_{C-C} ) h^2 / ( a l )$, where $d_{C-C} \sim a$ is the
length of the bond between nearest carbon atoms. In graphene, the
parameter $\beta = \partial \log ( t ) /
\partial \log ( d_{C-C} )$ is $\beta \approx 2$. Hence, we find that the flux
per ripple, in quantum units, is:
\begin{equation}
\Phi \approx \beta \frac{h^2}{l a}
\end{equation}
For $h \sim 1-5$nm and $l
\sim 50$nm, we find $\delta t / t \sim \delta \vf / \vf \sim 10^{-2} -
10^{-1}$, and $\Phi \gtrsim 1$.
\begin{figure}
\begin{center}
  \includegraphics*[width=3cm]{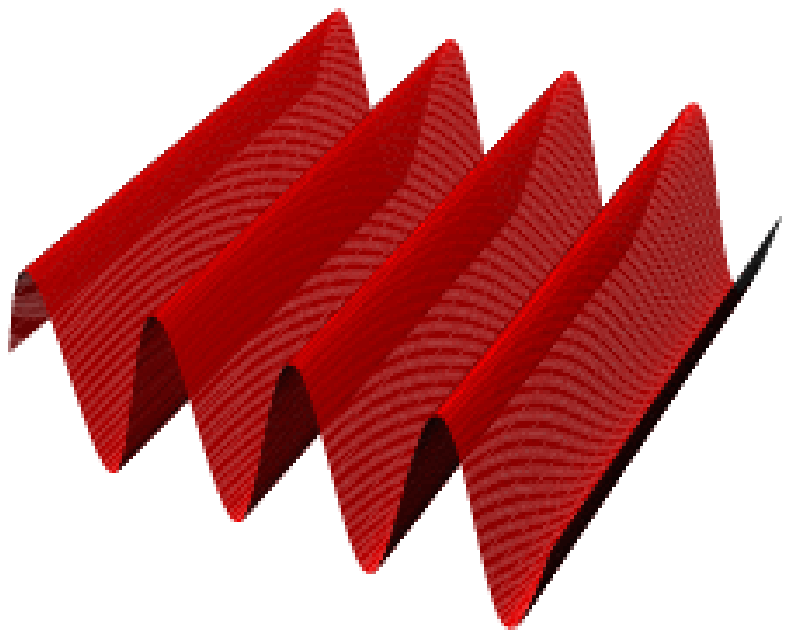}
\includegraphics*[width=3cm]{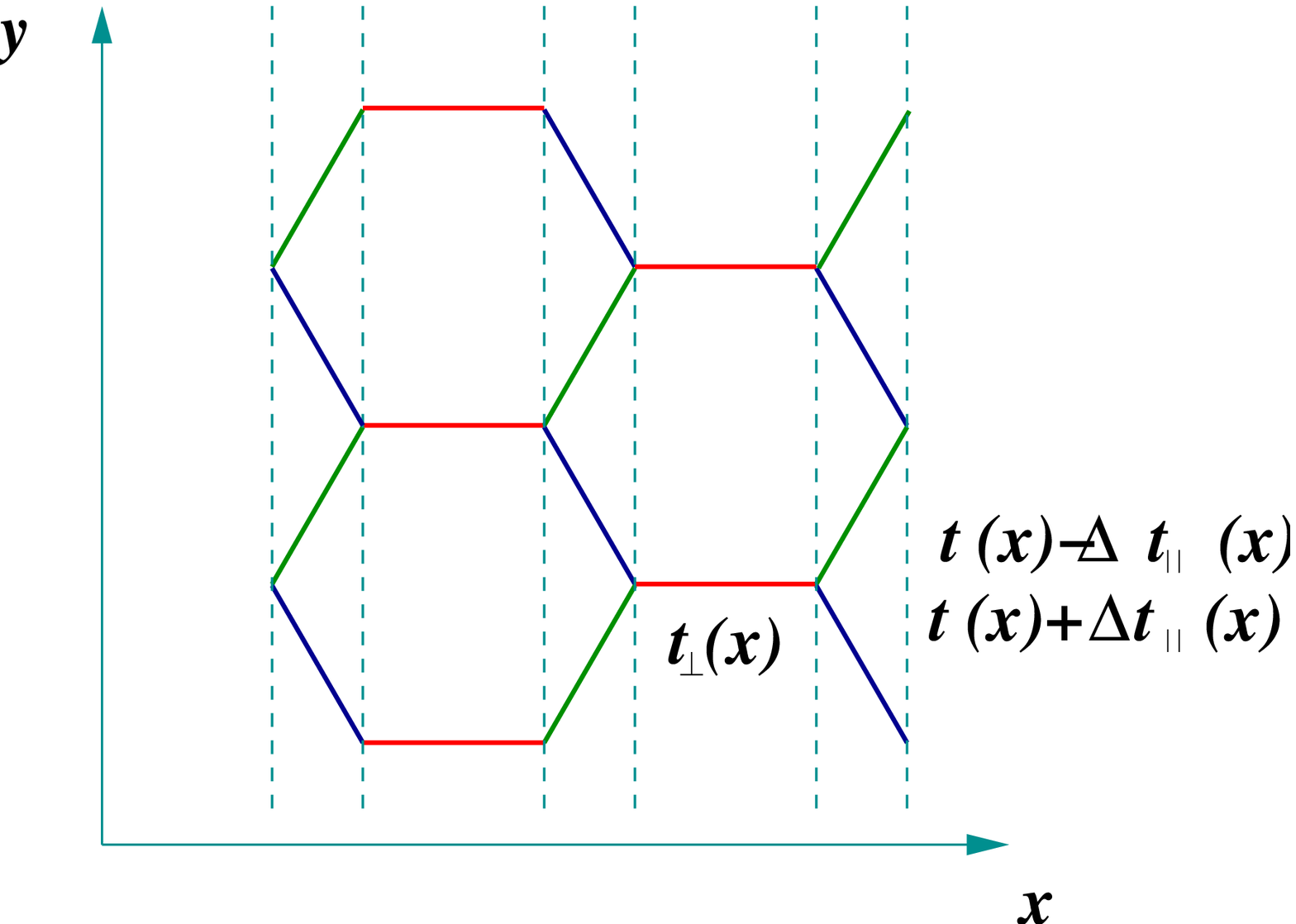}
\end{center}
\caption{(Color online). Left: sketch of the ripples considered in the text. Right:
  Modifications of the nearest neighbor hoppings induced by the ripple. See
  text for details.}
\label{sketch}
\end{figure}

\subsection{One dimensional ripples.}
We analyze numerically the emergence of midgap states as function
of the modulations of the hoppings using a simple model with the
geometry sketched in Fig.[\ref{sketch}].This translational symmetry along the
$y$ axis greatly simplifies the
calculations. The changes in the
electronic structure induced by the ripple are determined by the local value
of the gradient of
the tight binding hoppings. More complicated
patterns of ripples can be decomposed in regions described by an average
gradient of the hoppings. Hence, the model studied here should have similar
features to other structures, provided that the magnitude of the gradients of
the hoppings are comparable.

 The translational
invariance along the $y$ axis implies that $k_y$ is a good quantum
number. Hence, the hamiltonian can be reduced to a set of effective one
dimensional hamiltonians, one for each value of $k_y$, which
describe the hoppings between the rows shown on the right side of
Fig.[\ref{sketch}]. In the absence of modulations, the absolute
values of these hoppings are $t$ and $2 t \cos ( k_y \sqrt{3} a /
2 )$, were $a$ is the distance between carbon atoms. The low
energy states are centered around $k_y \sqrt{3} a = 2 \pi / 3$ and
$k_y \sqrt{3} a = 4 \pi / 3$. The modulation of the hoppings leads
to the replacement:
\begin{eqnarray}
t &\leftrightarrow &t_\parallel ( x ) \nonumber \\
2 t \cos \left( \phi \right)  &\leftrightarrow
 &\sqrt{
  \bar{t}_\perp^2 ( x ) \cos^2 \left( \phi \right) +
  \Delta t_\perp^2 ( x ) \sin^2 \left( \phi  \right)}
\label{hoppings}
\end{eqnarray}
where $\phi = ( \sqrt{3} k_y a ) / 2$. An external (real) magnetic field,
described by the vector
potential $A_y ( x ) = B x$ leads to the replacement of $k_y$ by
$k_y + e A_y / c$.

\begin{figure}
\begin{center}
\includegraphics*[width=4cm]{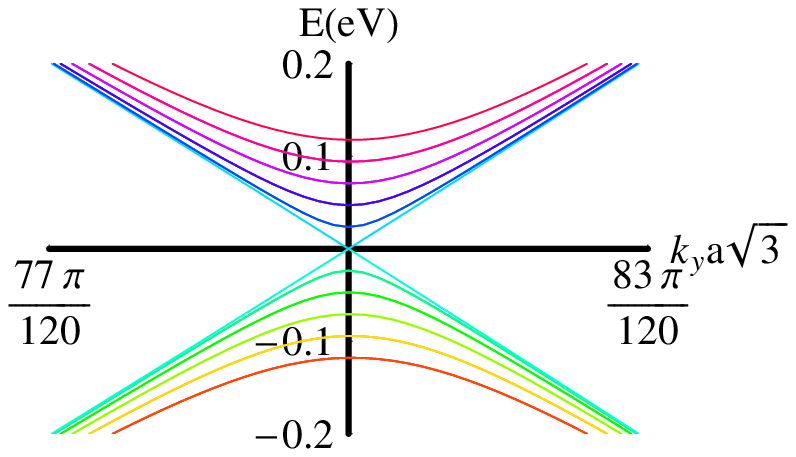}
\includegraphics*[width=4cm]{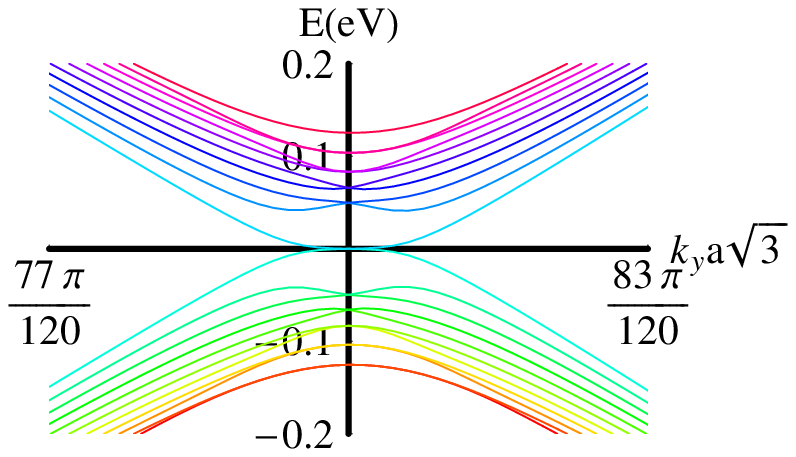}
\includegraphics*[width=4cm]{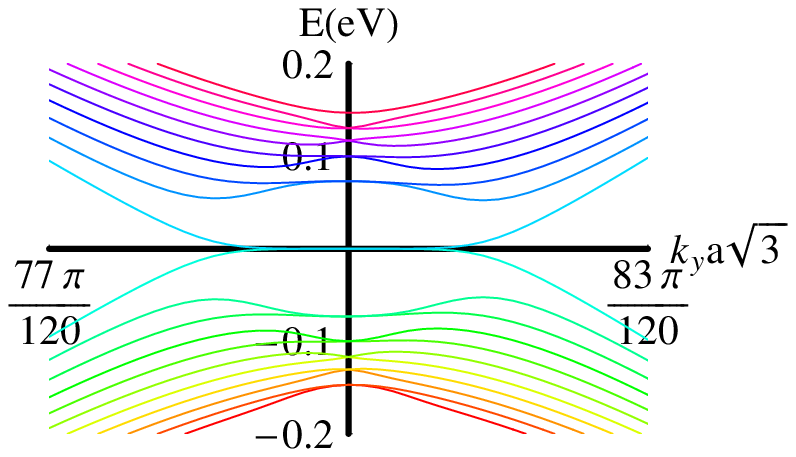}
\includegraphics*[width=4cm]{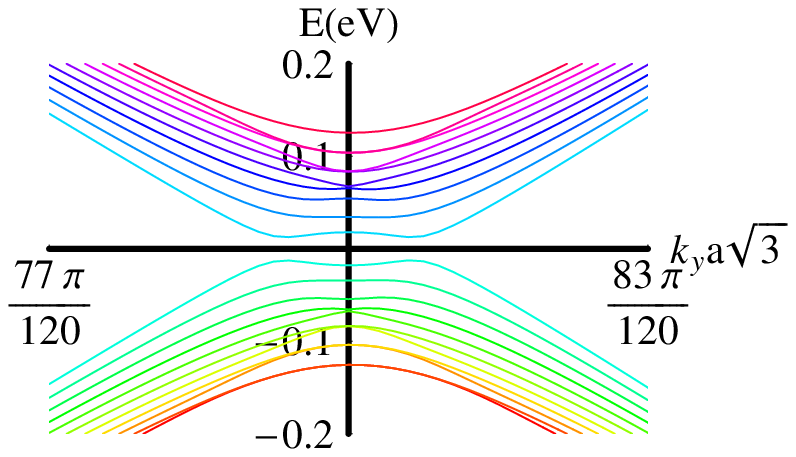}
\end{center}
\caption{(Color online). Low energy states induced by a ripple as shown in
  Fig.[\protect{\ref{sketch}}]. The average hopping is $t_\parallel =
  3$eV. The width of the
  ripple is $1200 a = 168$nm. The modulations of the hoppings are: Top left,
  $\delta t / t = 0$, top right, $\delta t / t = 0.02$, bottom left, $\delta t / t = 0.04$,
  bottom right, $\delta t / t = 0.02$, and a periodic electrostatic potential
  of amplitude 0.02eV.}
\label{bands_1}
\end{figure}

We have calculated the bands of the ribbons shown in Fig.[\ref{sketch}] using
the modulation:
\begin{equation}
t_\perp ( x ) = \delta t \sin \left( \frac{2 \pi x}{l} \right)
\label{modulation}
\end{equation}
Results are shown in Fig.[\ref{bands_1}], with $l = 1200 a$ (800
unit cells). Periodic boundary conditions are used. The plot
corresponds to one of the valleys. The levels in the two valleys, in the
absence of an external magnetic field, remain degenerate.

The calculations are consistent with the qualitative estimates made
above. The number of midgap states is proportional to the range of $k_y$
values where they are defined. The higher Landau levels are less well
defined, in agreement with the the fact that the description of a hopping
modulation as an effective gauge field becomes exact only at the Dirac
energy. The results are similar to those found in the analysis of the Quantum
Hall Effect in thick nanotubes in a perpendicular field\cite{PGGBO06}, where
the (real) magnetic field is modulated and has zero total flux. There is a
region in momentum space where the gaps between subbands have a
minimum, corresponding to the region in real space where the effective field
vanishes.

\begin{figure}
\includegraphics*[width=4cm]{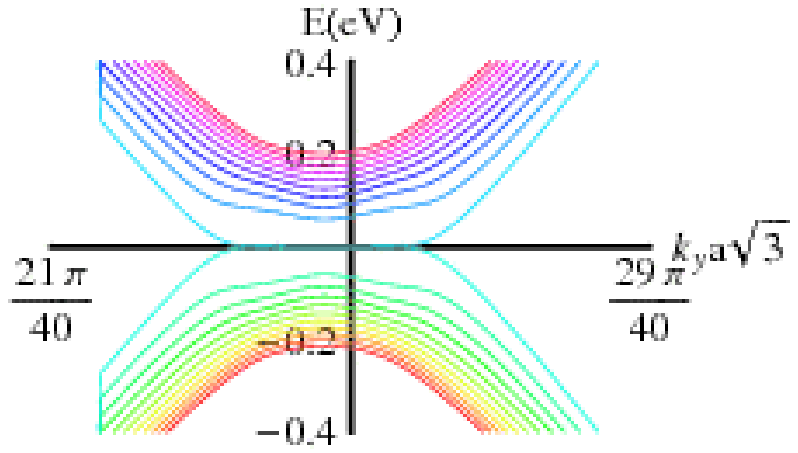}
\includegraphics*[width=4cm]{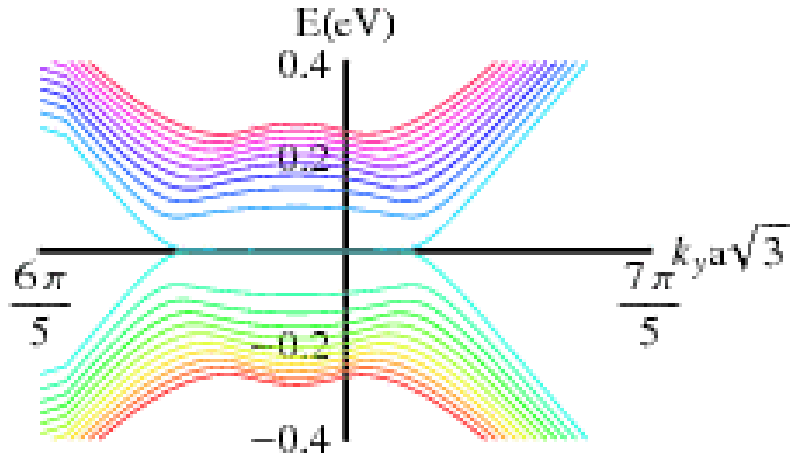}
\caption{(Color online). As in Fig.[\protect{\ref{bands_1}}], with a magnetic
  field $B = 10$
T. Left: $K$ valley. Right: $K'$ valley.} \label{bands_2}
\end{figure}

The effect of a uniform external magnetic field, which has the same
sign for the two valleys, is shown in Fig.[\ref{bands_2}]. The
combination of the breaking of inversion symmetry, induced by the
ripple, and time reversal symmetry, induced by the magnetic field,
leads to the inequivalence between the two valleys. The total
effective field is greater in one than in the other, as shown by the
larger region occupied by midgap states, and the fact that the
higher bands show less dispersion. The results presented in
Fig.[\ref{bands_2}] show that pseudomagnetic fields created by the
ripples broaden all Landau levels except the zero-energy one.
Interestingly, measurements of the activation gaps in the Quantum
Hall Effect regime in graphene demonstrate that zero-energy Landau
level is much narrower than the other ones \cite{GZKetal07}. The
results obtained using this model are in agreement with calculations
using the full valence band of graphene and the Local Density
Functional Approximation for one dimensional ripples of smaller
sizes\cite{WBKL07}.

\subsection{Two dimensional ripples.}
\begin{figure}
\begin{center}
\includegraphics*[width=3cm]{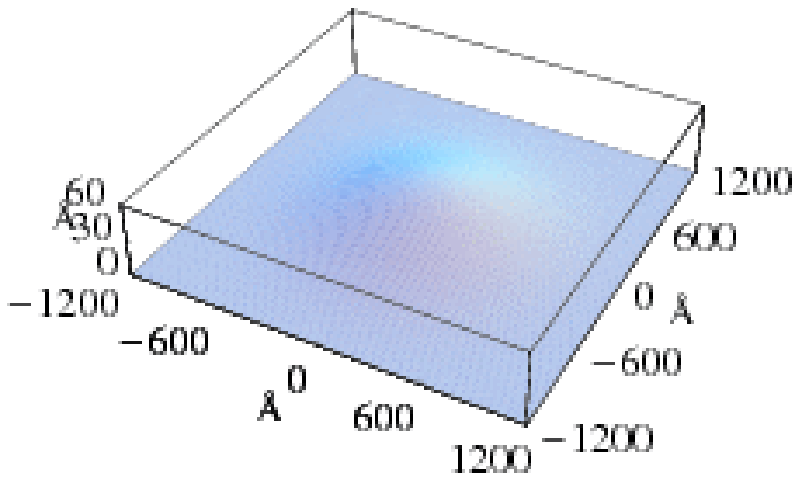}
\includegraphics*[width=3cm]{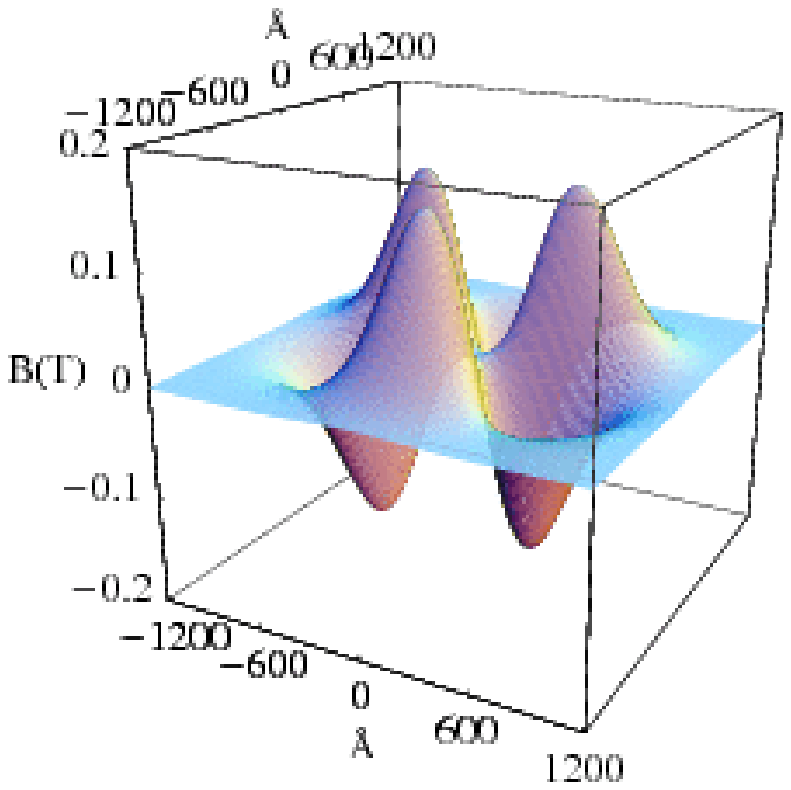}
\end{center}
\caption{(Color online). Right: Sketch of the ripple with an axially symmetric
  profile, similar to those  analyzed in the text. The length is $l = 600$\AA , and the
  height, is $h = 30$\AA. Left: Effective magnetic field generated by
  the ripple, see text for details.} \label{ripple_gaussian_sketch}
\end{figure}

We consider now a single two dimensional ripple with axial symmetry.
 The elastic
strains in circular coordinates are:
\begin{eqnarray}
u_{rr} &= &\partial_r u_r + \left( \partial_r h \right
)^2 \nonumber \\
u_{\theta \theta} &= &\frac{\partial_\theta u_\theta}{r^2} +
\frac{u_r}{r} + \frac{\left( \partial_\theta h \right)^2}{r^2}
\nonumber \\
u_{r \theta} &= &\partial_r u_\theta + \frac{\partial_\theta u_r}{r}
- \frac{u_\theta}{r} + \frac{ \partial_r h \partial_\theta h}{r}
\label{strain}
\end{eqnarray}
The Dirac equation in radial coordinates is:
\begin{widetext}
\begin{eqnarray}
i e^{i \theta} \vf \left( \partial_r + \frac{i \partial_\theta}{r}
\right) \Psi_A ( r , \theta ) + t \beta e^{- 2 i \theta} \left(
u_{rr} - u_{\theta \theta} - i u_{r \theta} \right) \Psi_A ( r ,
\theta ) &= &\epsilon \Psi_B ( r , \theta ) \nonumber \\
i e^{- i \theta} \vf \left( \partial_r - \frac{i \partial_\theta}{r}
\right) \Psi_B ( r , \theta ) + t \beta e^{ 2 i \theta} \left(
u_{rr} - u_{\theta \theta} + i u_{r \theta} \right) \Psi_B ( r ,
\theta ) &= &\epsilon \Psi_A ( r , \theta ) \label{dirac}
\end{eqnarray}
\end{widetext}
where $\beta = \partial \log ( t ) / \partial \log ( a )$, 
$t$ is the hopping between nearest neighbor orbitals, and $a$ is
the distance between carbon atoms.
If  the gauge field induced by the ripple has circular
symmetry, we obtain:
\begin{equation}
t \beta \left( u_{rr} -u_{\theta \theta} + i u_{r \theta} \right) =
f ( r ) \label{gauge}
\end{equation}
A sketch of the ripples studied here is shown in
Fig.[\ref{ripple_gaussian_sketch}], and also the associated effective
magnetic field. The parameters used in the figure, $l = 600$\AA \, and $h =
30$\AA \, give a flux of effective magnetic field of order unity per ripple.

We compute numerically the electronic levels of ripples with axial
symmetry, embedded in hexagons with periodic boundary conditions.
The hexagons are labeled by an integer $N$, such that the total
number of sites that they contain is $6 ( N*1)^2$. The side of the
hexagon is $L = \sqrt{3} ( N + 1 ) a$, where $a = 1.4$\AA \, is the
distance between carbon atoms. The strains are simulated by an in
plane deformation which modulates the hoppings:
\begin{equation}
t_{\vec{r}_i \vec{r}_j} = t_0 + t_0 \beta \frac{ \left( \delta
\vec{r}_i - \delta \vec{r}_j \right) \left( \vec{r}_i - \vec{r}_j
\right)}{a^2}
\end{equation}
with:
\begin{equation}
\delta \vec{r}_i =   g ( | \vec{r}_i | ) \frac{\vec{r}_i}{|
\vec{r}_i |}
\end{equation}
and:
\begin{equation}
t \beta g ( r ) = A  t r e^{-2 ( r / l )^2} \label{def_A}
\end{equation}
The resulting  strain tensor is the same, in the continuum limit, to
that is induced by a gaussian modulation of the height of the
graphene sheet, $h ( r ) = h_0 e^{-(r/l)^2}$, with $A = \beta ( h_0
/ l )^2$.

\begin{figure}
\includegraphics*[width=5cm]{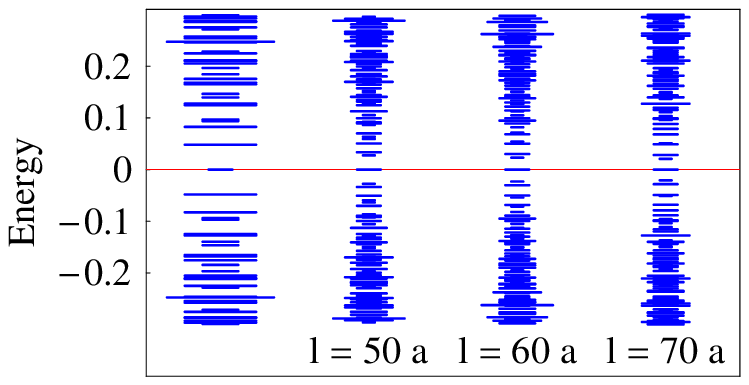}
\includegraphics*[width=5cm]{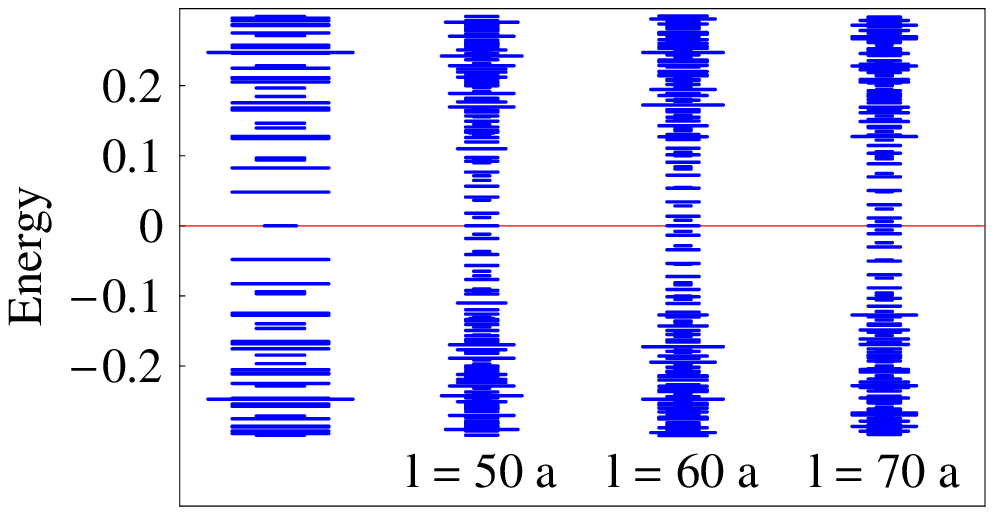}
\caption{(Color online). Energy levels (in units of $t$) for ripples
with circular symmetry embedded in an hexagon with $6 \times 52^2 =
16224$ atoms and periodic boundary conditions, and different radii
$l$ (in units of $a$). The length of each line is proportional to
the degeneracy of the level. Top: $A = 0.2$. Bottom: $A = 0.4$ (see
text for details).} \label{hexagon_levels}
\end{figure}

In the absence of the ripple, the levels have usually six- or
twelvefold degeneracy, and are separated by gaps of order $\vf / L$.
There is a fourfold degenerate state at $E=0$, as the number of unit
cells in the hexagon is a multiple of three. The electronic spectra
at low energies for hexagons with  $N = 76$, (the number of sites is
34656) with embedded ripples of different sizes are shown in
Fig.[\ref{hexagon_levels}]. The side of these hexagons is $L \approx
131 a \approx 18$nm. The ripple is defined as in eq.(\ref{def_A}),
with $A = 0.2$ and $A = 0.4$, and the radius of the ripple is $l =
50 a , 60 a$ and $70 a$. The modified hoppings range from 0.9$t$ to
1.2$t$ for $A = 0.2$, and between 0.8$t$ and 1.4$t$ for $A=0.4$.

\begin{figure}
\begin{center}
\includegraphics*[width=5cm]{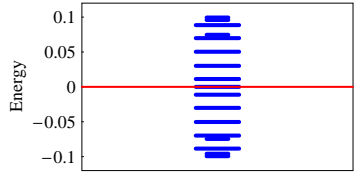}
\end{center}
\caption{(Color online). Low energy levels of the ripple with
$A=0.4$ and $l=70 a$ shown in Fig.[\protect{\ref{hexagon_levels}}].}
\label{hexagon_levels_amp}
\end{figure}
In the presence of ripples, the degeneracy of the levels is reduced,
and the number of low energy modes increases with the size of the
ripple, in agreement with the analytical estimates made above. Note
that there are still significant gaps between these states,
indicating that they will lead to peaks in the density of states,
and not to resonances within a broad continuum. 
The periodic
boundary conditions imply that the low energy levels of a given
ripple are hybridized with those in neighboring ones, leading to a
shift from $E=0$. The dependence of the level spacing with $\delta t$ and $l$ agrees with the
estimates in Section IIa. 

The low energy spectrum for a ripple of radius $l
= 70 a$ and amplitude $A= 0.4$ is shown in
Fig.[\ref{hexagon_levels_amp}]. The level spacing near $E=0$,
$\Delta E \approx 0.02 t$ implies an effective magnetic length, $l_B
=  \vf / \Delta E = ( 3 t a / 2 ) / \Delta E \approx 75 a \sim l$. This
length is larger, although the order of magnitude is comparable, than the
analytical estimate obtained in Section IIa, $l_B \sim \sqrt{(t/\delta t) (
  l a)} \sim \sqrt{70/0.3} \sim 15 a$. Given the uncertainities associated to
the estimate in Section IIa, and the fact that we have not taken into account
that the effective field induced by the ripple is split into six peaks of
smaller size than the ripple itself, we consider the agreement reasonable. 

\begin{figure}
\begin{center}
\includegraphics*[width=5cm]{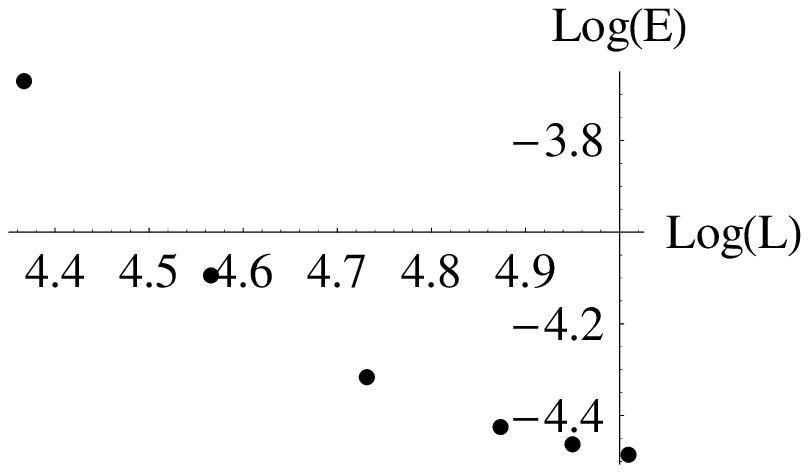}
\includegraphics*[width=5cm]{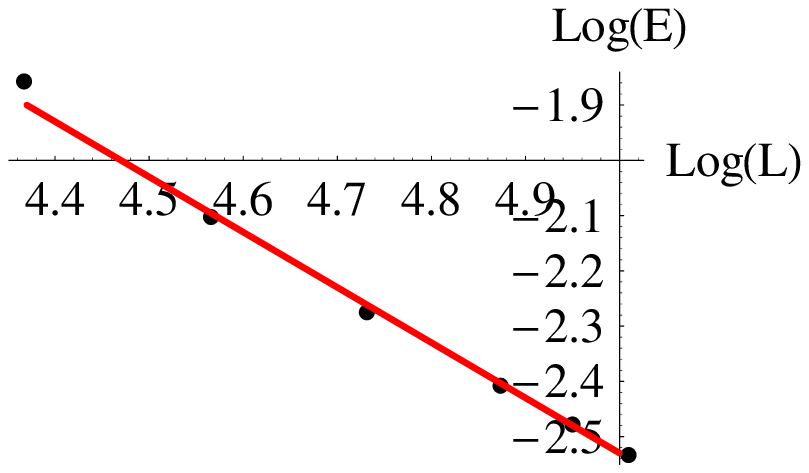}
\end{center}
\caption{(Color online). Top: Log-log plot of the scaling, as
function of the size of the hexagon, of the lowest energy state away
from $E=0$ of a system with a ripple of radius $l = 50 a$, and
amplitude $A=0.4$.
Bottom: Scaling of the fifth level away from $E=0$, and for the same
parameters. The red straight line has a slope of -1.}
\label{hexagon_levels_log}
\end{figure}
\begin{figure}
\begin{center}
\includegraphics*[width=5cm]{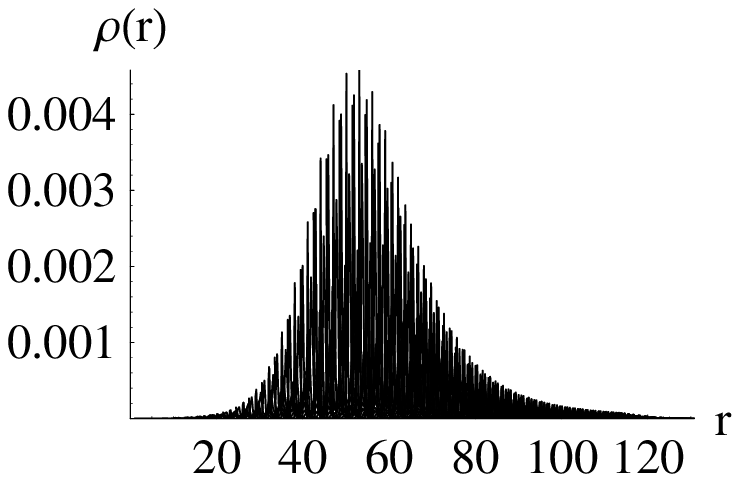}
\includegraphics*[width=5cm]{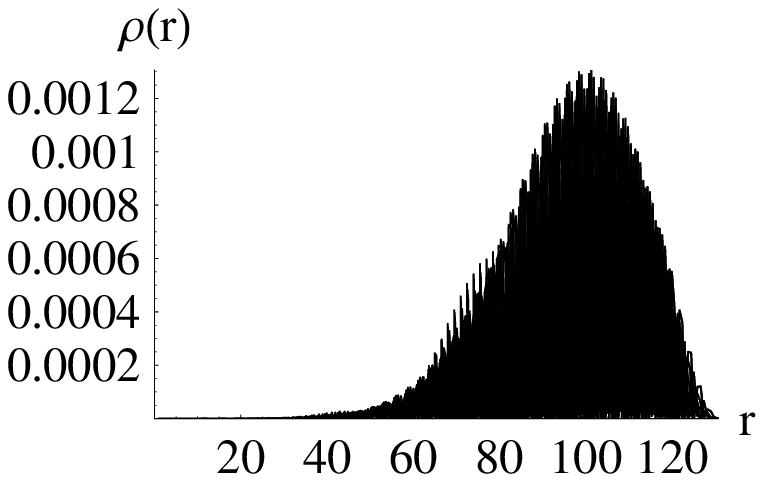}
\end{center}
\caption{(Color online). Top:  Density profile of the wavefunction
of the state closest to $E=0$, of a ripple with radius $l = 50 a$.
The energy of this state is shown in
Fig.[\protect{\ref{hexagon_levels_log}}]. Bottom: density for the
fifth level away from $E=0$.} \label{hexagon_wv}
\end{figure}

The scaling of the lowest level close to $E=0$ as function of system
size, $L$, for a fixed radius of the ripple, $l=50a$, is shown in
Fig.[\ref{hexagon_levels_log}]. The energy scales approximately as
$L^{-1}$ for $l \lesssim L$, and it shows a weaker dependence on $L$
for $l \ll L$. For comparison, the scaling of a level further away from $E=0$
shows the $L^{-1}$ scaling as function of system size expected from a
delocalized level described by the Dirac equation. The radial distribution of
the density associated to these wavefunctions is shown in
Fig.[\ref{hexagon_wv}]. In agreement with the scaling behavior of the energy,
one of the states is localized within the ripple, while the other is extended
towards the edges of the hexagon.

\begin{figure}
\begin{center}
\includegraphics*[width=5cm]{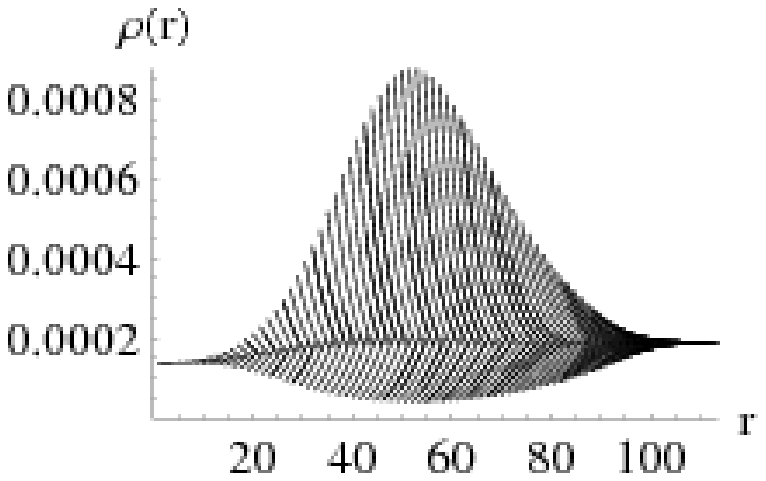}
\includegraphics*[width=5cm]{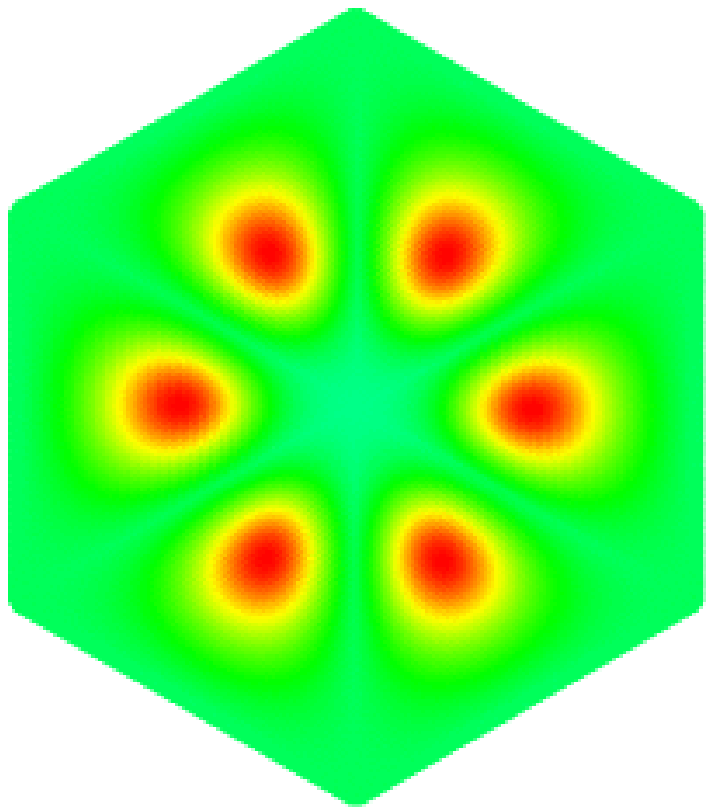}
\includegraphics*[width=5cm]{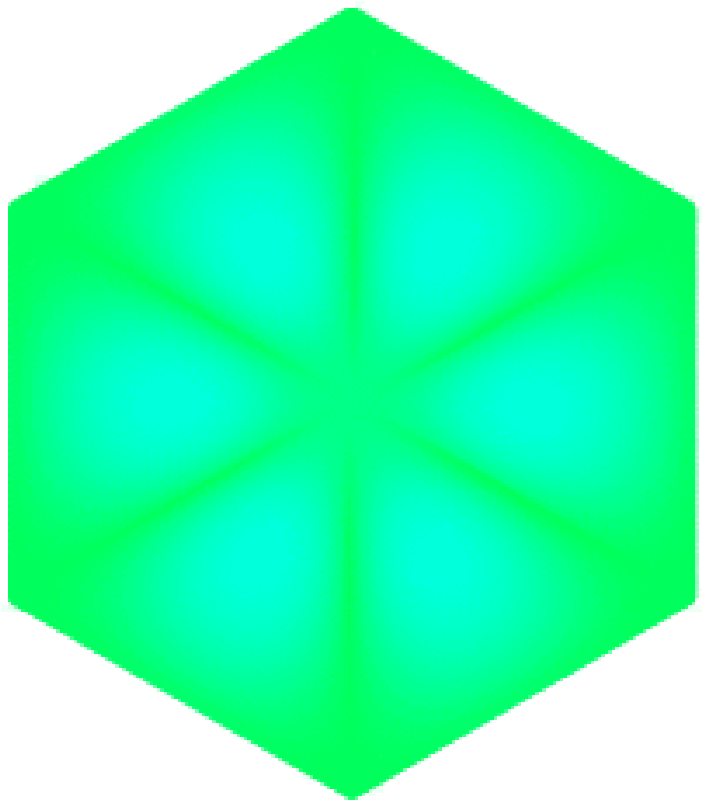}
\end{center}
\caption{(Color online). Top: Density as function of distance to the
center of the wavefunction of the state at $E=0$, of a ripple with
radius $l = 50 a$. Center: Density for sublattice A. Bottom: Density
for sublattice B. The color code is such that density decreases from
red to green, blue, and light blue.} \label{hexagon_zero}
\end{figure}

All the lattices studied, with and without the modulation of the hoppings,
show four zero energy states. In the clean system, the existence of these
states is determined by
the valley and sublattice degeneracy, and the density associated to them is
uniform throughout the lattice. The spatial extent of these states changes
qualitatively in the presence of a ripple, as shown in
Fig.[\ref{hexagon_zero}]. The wavefunctions become localized within the
ripple, leading to a peak in the local density of states at $E=0$, even if
these states are delocalized. The wavefunction shows the hexagonal symmetry
of the undelying lattice, and most of the charge is localized in six regions
of the ripple. At the neutrality point there are only two (four including
spin) electrons available in these states, opening the possibility of charge
fractionalization.

The results presented above have been obtained for large modulations
of the hoppings, or, alternatively, strong height corrugations. This
limitation is imposed by the dimension of the hamiltonian which
can be diagonalized.
 
The calculations show, however, that the low energy part
of the spectra have the features expected from the continuum Dirac
equation, and the wavefunctions associated to these states have a
smooth envelope on scales comparable to the lattice constant. Hence,
the method used can be considered a finite element technique which
approximates well the Dirac equation in the presence of a ripple. The
property of the lattice model which needs to be kept invariant is $\vf = ( 3
t a ) / 2$. 
The continuum equations are invariant
 under the scaling $l
\rightarrow \lambda l$, $\delta t \rightarrow \delta t / \lambda$,
and $E_n \rightarrow E_n / \lambda$. The scaling of $\delta t$ implies that
the corrugation of the ripple scales as $(h/l)^2 \rightarrow (h/l)^2 /
\lambda$. As a result, we can extrapolate the results
analyzed here to ripples of larger size and weaker corrugations. The
resulting effects are in agreement with the qualitative estimates made in
Section IIa.

\section{Interaction effects.} The analysis in the
preceding section shows that rippled graphene samples, in the
absence of interactions, have a peak in the density of states at
the Dirac energy. The width of this peak decreases as $e^{-
(\delta t / t ) ( l / a )}$, and the fraction of states, in an
area of size $l^2$, that it includes is proportional to the same
dimensionless combination, $f \sim (\delta t / t ) ( l /  a )$.
The resulting diverging electronic compressibility implies that
interactions will induce instabilities, in the same way as in
graphene in a magnetic field near half
filling\cite{GMSS06,GMD06,NM06,FB06,AF06,FL07,ANZLGL07,LS07}. As
shown in Fig.[\ref{bands_1}], while the midgap states are well
defined, the higher bands, for small values of $\delta t / t$,
tend quickly towards the clean graphene limit. If the interactions are weak,
this implies that their effect is limited to the states within a narrow
range of energies near the neutrality point. Alternatively, it can be argued
that interactions can be studied in a restricted hamiltonian which includes
only the midgap states. Hence, a reasonable
upper bound to the gap opened by many body effects is $\Delta \sim
\sqrt{( t \delta t ) ( a / l )}$.
\begin{figure}
\includegraphics*[width=4cm]{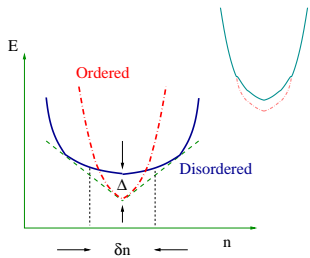}
\caption{(Color online). Sketch of the free energy of an ordered phase induced by
  interactions. The ordered phase is assumed to have a gap, and the
  derivative of the free energy at
  $n=0$ is discontinuous. On the other hand, the disordered phase is assumed
  to have a high electronic compressibility. The Maxwell construction
  indicated by the straight lines shows the region where electronic phase
  separation occurs. For comparison, an ordinary second order phase transition
  between phases with similar electronic compressibilities is also shown. See
  text for details.}
\label{phase_diagram}
\end{figure}

A  plausible ordered phase which can emerge at low temperatures is a
ferromagnetic phase\cite{AF06,NM06}. The saturation magnetization will be
small, $m \sim ( \delta t / t ) / ( a l )$. The Curie temperature needs not
be small, as transverse spin waves in a system like this do not induce a
large suppression of $T_C$\cite{EK06}. This phase may help to explain the
observation of ferromagnetism in graphite samples\cite{Eetal03}.

Phase transitions in systems where the electronic compressibility is large
are often of first order, and the system has a tendency towards electronic
phase separation in the ordered phase\cite{GGA02,S04}.
The expected
dependence of the free energy of the ordered and disordered phases as
function of electronic concentration is sketched in
Fig.[\ref{phase_diagram}].
The region of phase separation which can be obtained from a Maxwell
interpolation using the free energies of the different phases will
be replaced by a disordered phase with charge puddles, as the long
range Coulomb interaction suppresses macroscopic phase separation.
The resulting situation resembles that in magnetic systems with
striped phases or domains\cite{BHKSEH04,PRK06}.

We can make a qualitative estimate of the sizes of the charge
puddles by comparing the energy gain per unit area in the ordered
phase and the electrostatic cost of forming the puddle. For a puddle
of size $L_{pud}$, the energy in the ordered phase decreases by an
amount $\Delta n L_{pud}^2$, where $n$ is the density of electrons
contribute to the ordered phase. The electrostatic energy is $e^2
n^2 L_{pud}^3$. Then, the typical puddle size is $L_{pud} \sim
\Delta / ( e^2 n )$. Putting together the estimates made above, we
obtain:
\begin{eqnarray}
\Delta &= &\sqrt{\frac{t \delta t  a}{l}} \nonumber \\
n &= &\frac{\delta t}{t}\frac{1}{ l a} \nonumber \\
L_{pud} &= &l \frac{t a}{e^2} \sqrt{\frac{t}{\delta t} \frac{l}{a}}
\sim l \frac{\vf}{e^2} \sqrt{\frac{t}{\delta t} \frac{a}{l}}
\label{coulomb}
\end{eqnarray}
As in graphene $e^2 / \vf \sim 1$, and $\sqrt{\frac{t}{\delta t}
\frac{a}{l}} \sim O ( 1 )$ for reasonable ripple parameters, the
size of the puddle will be comparable with that of the ripple. A
sketch of the gap opened by an periodic electrostatic potential, $v ( x , y )
= v_0 \sin ( 2 \pi x / l )$, with $v_0 = 0.02$eV is shown in
the bottom right graph in Fig[\ref{bands_1}]. The gap is of the same order of
magnitude as the amplitude of the potential.

\section{Conclusions.}
We present analysis of the changes in the electronic structure of graphene
due to modulations in the hoppings induced by ripples and other sources of
elastic strains. The changes in the electronic structure are determined by
the dimensionless parameter $\Phi \sim ( \delta t / t ) ( l / a )  \sim (
\delta \vf / \vf ) ( l / a ) \sim ( \beta
h^2 ) / ( l a )$, where $\delta t / t
= \delta \vf / \vf$ is the modulation of the hopping parameter, or the Fermi
velocity, $\vf$, $h$ is the height of the rippls, $l$ is the size of the
ripple, and $a$ is the lattice constant. The 
parameter $\Phi$ gives the flux of effective magnetic field
threading an area of the size of the ripple. 

We find that reasonable values
for the size and height of a
ripple lead to the formation of midgap states.
These midgap states are similar to the Landau levels at the Dirac energy
induced by a magnetic field. The combination of hopping modulations and a
magnetic field breaks the symmetry between the two graphene valleys, leading
to the possibility of valley selection\cite{RTB07}, as electrons from
each valleys will scatter differently from extended defects\cite{CF06}.

Midgap states induce a large electronic compressibility when the
Fermi energy is near the Dirac point. Interaction effects will
lead to instabilities towards ordered phases, and electronic phase
separation, with typical puddle size not too different from that
of the ripple.

\section{Acknowledgments.} We thank L. Brey, P. le Doussal, A. Castro Neto, A. K. Geim,
J. Gonz\'alez, B. Horovitz, and K. S. Novoselov for useful
conversations. This work was supported by MEC (Spain) through grant
FIS2005-05478-C02-01, the Comunidad de Madrid, through the program
CITECNOMIK, CM2006-S-0505-ESP-0337, the European Union Contract
12881 (NEST)(F. G. and M. A. H. V.) and by  FOM (Netherlands) (M. K.
).
\section{Appendix. Analytical models of low energy states in rippled graphene.}
\subsection{Straight ripple.}
We analyze simple models of elastic deformations and the midgap states that
they may induce.

We study first a straight boundary between a stretched and a relaxed region
in graphene, as schematically shown in Fig.[\ref{ripple_flat}]. We describe
the change in the stretched region by a change in the nearest neighbor
hopping, $t$,  along the
horizontal direction, $t + \delta t$.

The Dirac hamiltonian in the stretched region, is:
\begin{equation}
{\cal H} \equiv \left( \begin{array}{cc} 0 & \vf ( -i k_x \mp k_y ) + \delta
    t\\ \vf (
    k_x \mp k_y + \delta t ) & 0 \end{array} \right)
\end{equation}
so that the perturbation induces a gauge field in the $y$ direction,
$A_y$. At the boundary, we have $\partial_x A_y \ne 0$.

The system has translational symmetry along the $y$ axis, so that
$k_y$ is conserved. The Dirac equation, for a given valley reads:
\begin{eqnarray}
\vf \partial_x \psi_A ( x ) \pm ( \vf k_y \mp \delta t ( x ) ) \psi_A ( x ) &=
&\epsilon \psi_B ( x ) \nonumber \\
- \vf \partial_x \psi_B ( x ) \pm ( \vf k_y \mp \delta t ( x ) ) \psi_B ( x )
&= &\epsilon \psi_A ( x )
\end{eqnarray}
These set of equations give the one dimensional Dirac equation with a gap
$\Delta ( x ) = \vf k_y + \delta t ( x )$. If the gap changes sign across the
interface they have localized solutions at $\epsilon = 0$. This condition
implies that either $\vf k_y > 0$ and $\vf k_y + \delta t < 0$ or $\vf k_y <
0$ and $\vf k_y + \delta t > 0$. Irrespective of the sign of $\delta t$,
there is a range of values of $k_y$:

\begin{figure}
\begin{center}
\includegraphics*[width=4cm]{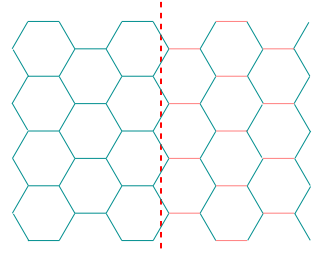}
\end{center}
\caption{(Color online). Sketch of a boundary between a stretched
and a relaxed region of
  graphene. }
\label{ripple_flat}
\end{figure}

\begin{equation}
\delta k_y = \frac{| \delta t |}{\vf}
\end{equation}
\begin{figure}[!]
\begin{center}
\includegraphics*[width=4cm]{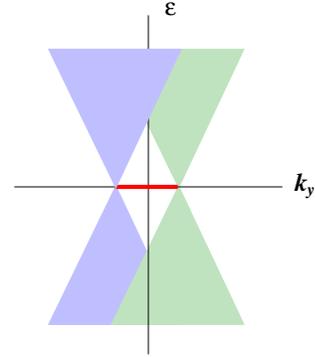}
\end{center}
\caption{(Color online). Electronic spectrum as function of $k_y$ of
the interface shown in
  Fig.[\protect{\ref{ripple_flat}}]. }
\label{bands_ripple}
\end{figure}
The electronic structure of the system is schematically shown in
Fig.[\ref{bands_ripple}]. The distortion shifts the Dirac cone in the
stretched region. A band of localized states joining the two Dirac cones is
induced. The number of midgap states per unit length of the ripple is:
\begin{equation}
n_{1D} = \frac{8 | \delta t |}{3 t a}
\end{equation}
where $t$ is the nearest neighbor hopping, and $a$ is the $C-C$ distance.

\subsection{Circular ripple.}
The zero energy wavefunction in a given valley,
in the presence of a circular ripple, obeys the Dirac equation,
eq.(\ref{dirac}):
\begin{equation}
i e^{i \theta} \vf \left( \partial_r + \frac{i \partial_\theta}{r} \right)
\Psi_A ( r , \theta ) + f ( r ) e^{- 2 i \theta} \Psi_A ( r , \theta ) = 0
\label{zero_dirac}
\end{equation}
and related equations for $\Psi_B ( r , \theta )$ and for the other valley.
This equation can be integrated analytically, and we obtain:
\begin{equation}
\Psi ( r , \theta ) = f ( r e^{i \theta} ) e^{i \frac{e^{- 3 i \theta}}{\vf
  r^3} \int_0^r
  d r' {r'}^3 f( r' )}
\end{equation}
where $f ( z )$ is an analytic function. This wavefunction is not
normalizable, irrespective of the choice of $f ( z )$. Choosing $f ( r e^{i
  \theta} ) = A$, where $A$ is a constant, the electronic density is:
\begin{equation}
\rho ( r , \theta ) = A^2 e^{\frac{2 \sin( 3 \theta)}{\vf r^3} \int_0^r
  d r' {r'}^3 f ( r' )}
\end{equation}
Assumming, for instance, that $\lim_{r \rightarrow \infty} f ( r ) \sim O
\left( e^{- 2 (r/l)^2} \right)$,and that $f ( r )$ does not diverge as $r
\rightarrow 0$, we find:
\begin{eqnarray}
\lim_{r \rightarrow 0} \rho ( r , \theta ) &= &A^2 \nonumber \\
 \lim_{r \rightarrow \infty} \rho ( r , \theta ) &= &A^2
\end{eqnarray}
The density has maxima at for a radial coordinate comparable to the
radius of the ripple, $r \sim l$, where it depends exponentially on
a quantity of order $f ( l ) l / \vf \sim \beta h^2 / ( l a )$. As a
function of $\theta$, $\rho ( r \sim l , \theta )$ has three maxima
and three minima.The positions of the maxima and the minima are
interchanged when analyzing the other valley. The resulting pattern
is in good agreement with the numerical results shown in
Fig.[\ref{hexagon_zero}].
\bibliography{ripples_v3}
\end{document}